\documentclass[aps,twocolumn,showpacs,superscriptaddress]{revtex4-1}
\usepackage{graphicx,color}
\usepackage{amsmath}
\usepackage{amssymb}
\usepackage{amsfonts}
\usepackage{bm}

\newcommand{\ot}{{\,\otimes\,}}
\newcommand{{\Cd}}{{\mathbb{C}^d}}

\def\oper{{\mathchoice{\rm 1\mskip-4mu l}{\rm 1\mskip-4mu l}%
{\rm 1\mskip-4.5mu l}{\rm 1\mskip-5mu l}}}
\def\<{\langle}
\def\>{\rangle}

\newtheorem{Example}{Example}
\newtheorem{Proposition}{Proposition}

\begin{document}

\title{Detecting non-Markovianity of quantum evolution via spectra of dynamical maps}
%Non-Markovian quantum dynamics: avoiding chronological product}
\author{Dariusz Chru\'sci\'nski$^1$, Chiara Macchiavello$^{2,3}$, and  Sabrina Maniscalco$^{4,5}$\\
%\affiliation{
$^1$Institute of Physics, Nicolaus Copernicus University, Faculty of Physics, Astronomy and Informatics, \\
Grudzi\c{a}dzka 5/7, 87--100 Toru\'n, Poland\\
$^2$ Quit group, Dipartimento di Fisica, Universit\'a di Pavia, via A. Bassi 6, I-27100 Pavia, Italy\\
$^3$ Istituto Nazionale di Fisica Nucleare, Gruppo IV, via A. Bassi 6, I-27100 Pavia, Italy \\
$^4$ Department of Physics and Astronomy, University of Turku, 20014 Turku, Finland \\
$^5$ Department of Applied Physics,  School of Science, Aalto University, P.O. Box 11000,
FIN-00076 Aalto, Finland}

%\title{Witnessing non-Markovianity of quantum evolution}
%\shorttitle{Witnessing non-Markovianity of quantum evolution} %Insert here a short version of the title if it exceeds 70 characters
%\author{Dariusz Chru\'sci\'nski \and Andrzej Kossakowski}
%\shortauthor{D. Chru\'sci\'nski \etal}
%\affiliation{Institute of Physics, Nicolaus Copernicus University \\
%Grudzi\c{a}dzka 5/7, 87--100 Toru\'n, Poland}
%
%\affiliation{Institute of Physics, Faculty of Physics, Astronomy and Informatics \\ Nicolaus Copernicus University \\
%Grudziadzka 5, 87--100 Torun, Poland }
%  \inst{2} Second Institute - Address
% }
%\pacs{03.65.Yz}{Decoherence; open systems; quantum statistical methods}
%\pacs{03.65.-w}{Quantum mechanics}
%\pacs{42.50.Lc}{Quantum fluctuations, quantum noise, and quantum jumps}
%\pacs{02.30.Tb}{Operator theory}

\begin{abstract}
We provide an analysis on non-Markovian quantum evolution based on the spectral properties of dynamical maps.
{We introduce the dynamical analog of entanglement witness to detect non-Markovianity and we illustrate its behaviour with several instructive examples. It is shown that for a certain class of dynamical maps the  shape of the body of accessible states provides a simple non-Markovianity witness. }
%In particular the dynamical analog of entanglement witness is provide which may be used as a simple tool for detecting non-Markovianity.
%Our analysis is illustrated by several instructive examples.
\end{abstract}

%\begin{document}

\pacs{03.65.Yz, 03.65.Ta, 42.50.Lc}

\maketitle

{\em Introduction} --- Spectral Theorem is one of the mathematical pillars of
quantum theory \cite{VN}. This celebrated result of von Neumann states that
for any normal operator  (i.e  $AA^\dagger = A^\dagger A$) in the Hilbert
space one has the corresponding spectral decomposition $A = \sum_k a_k
|\phi_k\>\<\phi_k|$ with complex $a_k$ and $\<\phi_k|\phi_l\>=\delta_{kl}$.
In particular if $A$ is not only normal but also Hermitian then $a_k$ are real. Many problems from quantum physics are directly related to finding the spectrum $\{ a_k\}$ of some normal/Hermitian operator.

In this Letter we apply some tools from spectral analysis to study the
evolution of an open quantum system. Such systems provide a fundamental tool
to study the interaction between a quantum  system and its environment,
causing dissipation, decay, and
decoherence \cite{Breuer,Weiss,Alicki}. It is, therefore, clear that open
quantum systems are important
for quantum-enhanced applications, as both entanglement and
quantum coherence are basic resources in modern quantum
technologies, such as quantum communication, cryptography,
and computation \cite{QIT}.

Recently, much effort {has been} devoted to {the} description, analysis and classification of non-Markovian quantum evolution (see e.g.
recent review papers \cite{rev1,rev2}). In analogy to entanglement theory
\cite{HHHH}  several  non-Markovianity measures were proposed which
characterize various concepts of non-Markovianity. The two most influential
approaches to non-Markovian evolution are based on divisibility  of dynamical
maps \cite{Wolf-Isert,RHP} and distinguishability of states  \cite{BLP} (for
other approaches see also
\cite{Fisher,fidelity,Luo1,Luo2,Bogna,Pater,Adesso}). The results we present
in this Letter allow to introduce for the first time a witness of
non-Markovianity in the same spirit of entanglement witnesses.
An entanglement witness method applied to the Choi-Jamiolkovski state
of a quantum channel was recently developed \cite{MR} in order to detect
properties based on convexity features. The method was tested experimentally
for entanglement breaking channels and for separable random unitary channels
\cite{exp-qcd}.

Besides the
fundamental interest, our approach simplifies, in certain cases, the
experimental detection of non-Markovianity of a dynamical map.

Let us recall that a dynamical map $\Lambda_t$ is CP-divisible if for any $t > s$ one has
%\begin{equation}\label{}
 $ \Lambda_t = V_{t,s} \Lambda_s$,
%\end{equation}
with $V_{t,s}$ being completely positive. We call quantum evolution Markovian iff the corresponding dynamical map is CP-divisible. Recently, this notion was refined as follows \cite{PRL-Sabrina}: $\Lambda_t$ is $k$-divisible iff $V_{t,s}$ is $k$-positive. In particular 1-divisible maps are called P-divisible ($V_{t,s}$ is positive). Maps which are even not P-divisible were called {\em essentially non-Markovian}.  { These types of dynamical maps have been recently simulated and detected experimentally \cite{test}}.

Note that CP-divisibility is {a} mathematical property of the map. Another approach more operationally oriented is based on distinguishability of quantum states  \cite{BLP}: we call quantum evolution BLP-Markovian if
\begin{equation}\label{BLP}
  \frac{d}{dt} ||\Lambda_t[\rho_1 - \rho_2]||_1 \leq 0 ,
\end{equation}
for any pair of initial states $\rho_1$ and $\rho_2$. Actually, assuming that $\Lambda_t$ is invertible one shows \cite{PRL-Sabrina}  that $\Lambda_t$ is $k$-divisible iff
%\begin{equation}\label{}
 $ \frac{d}{dt} || (\oper_k \ot \Lambda_t)[X]||_1 \leq 0$,
%\end{equation}
for all Hermitian $X \in M_k(\mathbb{C}) \ot \mathcal{B}(\mathcal{H})$. Note, that if $k=1$ and $X = \rho_1 - \rho_2$ one recovers (\ref{BLP}).

In {the following} we develop further the analysis of non-Markovian evolution
based on the spectral properties of dynamical maps, and provide
the dynamical analog of entanglement witness for detecting non-Markovianity.
{ Our} analysis is restricted to a class of commutative dynamical maps.
However,  the majority of well known examples {of open systems dynamics
belongs to} this class of maps.

\vspace{.4cm}

{\em Volume and body of accessible states   ---} Let us denote by $B$ the space of density operators. Clearly $B(t) = \Lambda_t[B]$ denotes the body of accessible states at time $t$. In a recent paper \cite{Pater} an interesting geometric characterization {is proposed}, namely, if $\Lambda_t$ is P-divisible, then
\begin{equation}\label{Vol}
  \frac{d}{dt} {\rm Vol}(t) \leq 0 ,
\end{equation}
where ${\rm Vol}(t)$ denotes the volume of accessible states at time $t$, i.e the volume of the convex body $B(t)$. This result follows from the fact that ${\rm Vol}(t) = |{\rm Det}\Lambda_t| {\rm Vol}(0)$ and for P-divisible map one has $\frac{d}{dt}  |{\rm Det}\Lambda_t| \leq 0$ (cf. \cite{WOLF}).

Let us provide more geometrical insight passing  to the  matrix representation $\Lambda_t \rightarrow F_{\alpha\beta}(t) := {\rm Tr}(G_\alpha \Lambda_t[G_\beta])$, where $G_\alpha$ is an orthonormal basis in $\mathcal{B}(\mathcal{H})$.   A suitable choice of $G_\alpha$ is the set of generalized Gell-Mann matrices with $G_0 = \mathbb{I}/\sqrt{d}$ and Hermitian $G_\alpha$ ($\alpha=1,\ldots,d^2-1$). In this case $F(t)$  has the following form
\begin{equation}\label{F(t)}
  F(t) = \left( \begin{array}{c|c} 1 & 0 \\ \hline \mathbf{q}_t & \Delta_t \end{array} \right) \ ,
\end{equation}
with $\mathbf{q_t} \in \mathbb{R}^{d^2-1}$ and $\Delta_t$ being $(d^2-1)\times (d^2-1)$ real matrix. It is clear that $F(t)$ encodes all properties of the original dynamical map $\Lambda_t$. In particular $\Lambda_t$ and $F(t)$ have exactly the same spectrum $\lambda_\alpha(t)$  ($\alpha=0,1,\ldots,d^2-1$, where $d = {\rm dim}\mathcal{H}$), and hence ${\rm Det}\Lambda_t = {\rm Det}F(t)= {\rm Det}\Delta_t$. This shows that the volume of the set of accessible states is fully controlled by the matrix $\Delta_t$ itself. Now, defining the generalized Bloch representation $\rho = \frac 1d (\mathbb{I} + \sum_{\alpha=1}^{d^2-1} x_\alpha G_\alpha)$, the action of the channel $\Lambda_t$ on $\rho$ corresponds to the following affine transformation of the generalized Bloch vector $\mathbf{x} \rightarrow \mathbf{x}_t = \Delta_t \mathbf{x} + \mathbf{q}_t$. If $\mathbf{x}_1$ and $\mathbf{x}_2$ are Bloch vectors corresponding to $\rho_1$ and $\rho_2$, then $[\rho_1-\rho_2] \rightarrow \Lambda_t[\rho_1 - \rho_2]$ corresponds to linear transformation $\Delta_t(\mathbf{x}_1-\mathbf{x}_2)$ and hence does not depend upon the vector $\mathbf{q}_t$. It clearly shows that BLP-Markovianity is controlled only by $\Delta_t$ whereas the full P-divisibility by the entire map $F(t)$, i.e. both $\Delta_t$ and $\mathbf{q}_t$. Note that divisibility of $F(t)$, that is, $F(t) = F(t,s)F(s)$ implies quite nontrivial relations $  \Delta_t = \Delta_{t,s} \Delta_s$ and $\mathbf{q}_t = \mathbf{q}_{t,s} + \Delta_{t,s}\mathbf{q}_s$, where $\mathbf{q}_{t,s}$ and $\Delta_{t,s}$ parameterize $F(t,s)$. They considerably simplify if the dynamical map $\Lambda_t$ is unital. In this case $\mathbf{q}_t=0$ and one is left with a simple divisibility condition $ \Delta_t = \Delta_{t,s} \Delta_s$.

\begin{Proposition} If $\Lambda_t$ is P-divisible and unital, then
\begin{equation}\label{}
   \frac{d}{dt} ||\Lambda_t[X] ||_2 \leq 0 ,
\end{equation}
for all normal operators $X$.
\end{Proposition}
For the proof see \cite{SM}.  In particular $ \frac{d}{dt} ||\Delta_t \mathbf{x} ||_2 \leq 0 $ which shows that the Euclidean norm of the Bloch vector $\mathbf{x}$ decreases monotonically \cite{SM}.

It should be clear that the volume of accessible states provides {a rather} weak witness -- one may easily construct maps satisfying (\ref{Vol}) which are not P-divisible. In particular it says nothing about the shape of the body of accessible states. For example very often during the evolution of a qubit the initial Bloch ball is deformed to an ellipsoid. P-divisible dynamics always decreases its volume but what about the length of the corresponding axis? Could one relate P-divisible evolution to the shape of accessible states? Moreover, it should be stressed that the shape is controlled by singular values of $F(t)$ and not by the spectrum itself. To analyze this problem let us consider singular value decomposition of the matrix $F(t)$
\begin{equation}\label{OO}
  F(t) = \mathcal{O}_1(t) \Sigma(t) \mathcal{O}^{-1}_2(t) ,
\end{equation}
where $\mathcal{O}_k(t)$ ($k=1,2$) are orthogonal matrices and $\Sigma(t)$ is a diagonal matrix containing singular values of $F(t)$. Hence the action of $F(t)$ consists in a rotation $\mathcal{O}^{-1}_2(t)$, a contraction governed by $\Sigma(t)$  (all singular values $\sigma_k(t) \leq 1$) followed by the rotation $\mathcal{O}_1(t)$. Since rotation does not change the volume {the latter} is fully controlled by $\Sigma(t)$. However, the shape of $B(t)$ depends both upon $\mathcal{O}_1(t)$ and $\mathcal{O}_2(t)$.

\vspace{.4cm}

{\em Commutative maps} --- %In this Letter we restrict to commutative maps,
%%%%%%%
%
To provide a stronger witness we restrict {our analysis to} a class of quantum evolutions  satisfying the following commutativity condition
\begin{equation}\label{COM}
  \Lambda_t \Lambda_s = \Lambda_s \Lambda_t ,
\end{equation}
for any $t,s >0$.  Equivalently, the time-local generator satisfies $\mathcal{L}_t \mathcal{L}_s = \mathcal{L}_s \mathcal{L}_t$. Commutativity condition (\ref{COM}) implies that $\Lambda_t$ and its dual (Heisenberg picture) possess time independent eigenvectors
\begin{equation}\label{COM-1}
  \Lambda_t[X_\alpha] = \lambda_\alpha(t) X_\alpha \ , \ \   \Lambda^*_t[Y_\alpha] = {\lambda^*_\alpha}(t) Y_\alpha ,
\end{equation}
for $\alpha = 0,1,\ldots,d^2-1$. This {condition} is {indeed} very restrictive. However, in practice the majority of the examples considered in the literature belong to the commutative class. The reason is very simple: assuming that $\Lambda_t$ satisfies the time-local master equation
%\begin{equation}\label{}
 $ \frac{d}{dt} \Lambda_t = \mathcal{L}_t \Lambda_t$,
%\end{equation}
with suitable time-local generator $\mathcal{L}_t$, one has
%\begin{equation}\label{T}
 $ \Lambda_t = \mathcal{T} e^{\int_0^t \mathcal{L}_\tau d\tau}$,
%\end{equation}
where $\mathcal{T}$ denotes the chronological operator. In  general the above formula has only a formal meaning and it is defined by the Dyson expansion
$\Lambda_t = \oper + \int_0^t dt_1 \mathcal{L}_{t_1} + \int_0^t dt_1 \int_0^{t_1} dt_2 \mathcal{L}_{t_1} \mathcal{L}_{t_2} + \ldots$.
Now, in the commutative case (\ref{COM}) the chronological product drops out and the solution is represented by the simple exponential formula $\Lambda_t = e^{\int_0^t \mathcal{L}_\tau d\tau}$. Moreover, %having found the damping bases $\{ F_\alpha\}$ and  $\{G_\alpha\}$ with ${\rm tr} (F_\alpha^\dagger G_\beta) = \delta_{\alpha\beta}$
the eigenvalues $\lambda_\alpha(t)$ of the dynamical map are related to the corresponding eigenvalues $\mu_\alpha(t)$ of the time-local generator $L_t$ via $\lambda_\alpha(t) = e^{\int_0^t \mu_\alpha(\tau)d\tau}$. One has, therefore, {the following} obvious {property}:
\begin{Proposition} If $\Lambda_t$ defines commutative P-divisible map, then
\begin{equation}\label{l}
   \frac{d}{dt} |\lambda_\alpha(t)| \leq 0 ,
\end{equation}
or equivalently ${\rm Re}\,\mu_\alpha(t) \leq 0$ for $\alpha=1,\ldots,d^2-1$.
\end{Proposition}
It is clear that the set of inequalities (\ref{l}) is much more restrictive that the single condition (\ref{Vol}) which immediately follows form (\ref{l}) due to ${\rm Det} F(t) = |\lambda_1(t) \ldots \lambda_{d^2-1}(t)|$.
Let us observe that for commutative maps condition (\ref{Vol}) may be easily translated to the condition upon the time-local generator $\mathcal{L}_t$. Using the well known property of matrices ${\rm Det}\, e^A = e^{{\rm Tr}A}$ one finds that (\ref{Vol}) is equivalent to
\begin{equation}\label{}
  {\rm Tr} \mathcal{L}_t \leq 0 ,
\end{equation}
where {with} ${\rm Tr}\mathcal{L}_t$ we mean the sum of eigenvalues or equivalently the trace of the matrix $L(t)$ defined by $L_{\alpha\beta}(t) = {\rm Tr}(G_\alpha \mathcal{L}_t[G_\beta])$.
In entanglement theory one defines an entanglement witness, i.e. an Hermitian operator $W$ in $\mathcal{H} \ot \mathcal{H}$ such that: $i)$ $\< \psi_1 \ot \psi_1 | W |\psi_1 \ot \psi_2\> \geq 0$, and $ii)$ ${\rm Tr}(W \rho) < 0 $ for some entangled state $\rho$. Any such operator may {be} constructed {as} $W := (\oper \ot \Phi)|\alpha\>\<\alpha|$, where $\Phi : \mathcal{B}(\mathcal{H}) \rightarrow \mathcal{B}(\mathcal{H})$ is a positive but not completely positive map, and
\begin{equation}\label{}
  |\alpha\> = \frac{1}{\sqrt{d}} \sum_{i=1}^d |i \ot i\> ,
\end{equation}
denotes the maximally entangled state in $\mathcal{H} \ot \mathcal{H}$. Consider now an arbitrary linear map $\Phi : \mathcal{B}(\mathcal{H}) \rightarrow \mathcal{B}(\mathcal{H})$ and define
\begin{equation}\label{}
  f_\Phi = \<\alpha|(\oper \ot \Phi)[P^+]|\alpha\> \ , % = \<\alpha|W_\Phi|\alpha\>
\end{equation}
with $P^+ = |\alpha\>\<\alpha|$.
%Note, that $(\oper \ot \Phi)|\alpha\>\<\alpha|$ is a positive operator and hence $f_\mathcal{E} \geq 0$.
Interestingly, $f_\Phi$ is fully characterized by the spectral properties of the map $\Phi$.
One has the following

\begin{Proposition} Function $f_\Phi$ is fully determined by the spectrum of $\Phi$, that is, 
%\begin{equation}\label{}
 $ f_\Phi = d^{-2} \sum_{\alpha=0}^{d^2-1} \lambda_\alpha $,
%\end{equation}
where $\lambda_\alpha$ are eigenvalues of $\Phi$.
\end{Proposition}
Indeed, consider the spectral representation
%\begin{equation}\label{}
 $ \Phi[\rho] = \sum_\alpha \lambda_\alpha F_\alpha {\rm Tr}(G^\dagger_\alpha \rho)$,
%\end{equation}
where $\{F_\alpha,G_\alpha\}$ provide a damping basis \cite{damping} for the map $\Phi$, that is, $\Phi[F_\alpha] = \lambda_\alpha F_\alpha$ and $\Phi^*[G_\alpha] = \lambda^*_\alpha G_\alpha$ such that ${\rm Tr}(F_\alpha G^\dagger_\beta)=\delta_{\alpha\beta}$. {One has}
%Of course if the channel  is hermitian, then $\lambda_\alpha$ is real and $F_\alpha=G_\alpha$.
\begin{eqnarray*}
  d^2 f_\Phi &=&  \sum_{i,j} \sum_{k,l} {\rm Tr}( (|i\>\<j| \ot |i\>\<j|) \cdot (|k\>\<l| \ot \Phi[  |k\>\<l|]))  \\
  &=& \sum_\alpha \sum_{i,j} \lambda_\alpha \<i|F_\alpha|j\> \< j| G^\dagger_\alpha| i\> =  \sum_\alpha  \lambda_\alpha,
 %  &=& 
\end{eqnarray*}
due to ${\rm Tr}(F_\alpha G^\dagger_\alpha)=1$. {If} the corresponding dynamical map $\Lambda_t = \exp( \int_0^t \mathcal{L}_\tau d\tau )$ is commutative then (\ref{Vol}) is equivalent to
\begin{equation}\label{EW-L}
  \<\alpha|(\oper \ot \mathcal{L}_t)[P^+]|\alpha\> \leq 0 \ .
\end{equation}
Hence, the violation of (\ref{EW-L}) may be considered as {\em dynamical analog of an entanglement witness}.

%Let us observe that the dynamical map $\Lambda_t$ is commutative if the corresponding family of matrices  $A(t)$ is commutative, i.e.
%\begin{equation}\label{Delta-q}
%\Delta_t \Delta_s = \Delta_s \Delta_t\ , \ \ [\Delta_t - \mathbb{I}]{\bf q}_s = [\Delta_s -\mathbb{I}]{\bf q}_t .
%\end{equation}

% We stress that this result holds for any P-divisible map. Now,
%$$ |{\rm Det} F(t)| = |\lambda_1(t) \ldots \lambda_{d^2-1}(t)| , $$
%and hence if the map is commutative then $\frac{d}{dt} |{\rm Det} F(t)| \leq 0$ follows from (\ref{l}). It is clear that the set of inequalities (\ref{l}) is much more restrictive that the single condition (\ref{Vol}). One may pose a natural question: could we characterize a shape of the body of states at time $t$? How the shape evolves at time?

\vspace{.4cm}

{\em Normal commutative maps} --- Consider a class of commutative maps (\ref{COM}) which are normal
\begin{equation}\label{NOR}
   \Lambda_t \Lambda_t^* = \Lambda_t^* \Lambda_t ,
\end{equation}
for any $t \geq 0$. This condition guarantees that due to the spectral theorem both $\Lambda_t$ and its dual $\Lambda^*_t$ (Heisenberg picture) possess common eigenvectors, i.e. $X_\alpha = Y_\alpha$. Moreover, since $\Lambda_t^*$ is unital, i.e. $\Lambda_t^*[\mathbb{I}]=\mathbb{I}$, the map $\Lambda_t$ is unital as well.  One may {choose} therefore $X_0 = \mathbb{I}/\sqrt{d}$, with $d = {\rm dim}\mathcal{H}$, which corresponds to $\lambda_0(t)=1$. Using the matrix representation (\ref{F(t)}) it means that $\mathbf{q}_t=0$ and $\Delta_t$ is a normal matrix.  It is well known that the shape of the body of states is not controlled by eigenvalues $\lambda_\alpha(t)$ but by singular values $s_\alpha(t)$. Note, however, that if the map is normal, then
%\begin{equation}\label{}
 $ s_\alpha(t) = |\lambda_\alpha(t)|$
%\end{equation}
and hence conditions (\ref{l}) are equivalent to
%\begin{equation}\label{s}
 $ \frac{d}{dt} s_\alpha(t) \leq 0$.
%\end{equation}
Let us observe that the formula (\ref{OO}) implies the following 
%body of states $B(t)$ at time $t$ satisfies $B(t) = \Delta_t[B(0)] \subset \mathbb{R}^{d^2-1}$.

\begin{Proposition}
If $\Lambda_t$ is a P-divisible commutative normal dynamical map, then there exists a family of orthogonal matrices $\mathcal{O}(t,s) \in O(d^2-1)$ such that
\begin{equation}\label{}
  \mathcal{O}(t,s)[B(t)] \subset B(s) ,
\end{equation}
for all $t \geq s$.
\end{Proposition}
The role of $\mathcal{O}(t,s)$ is to rotate $B(t)$ with respect to $B(s)$ such that $\mathcal{O}(t,s)[B(t)]$ is contained within $B(s)$.
Hence, for P-divisible commutative normal maps $\Lambda_t$, not only ${\rm Vol}(t)$ decreases in time but also the body itself $B(t)$ (up to orthogonal rotation) shrinks in time.

%It is well known that for unital channels one has $||\mathcal{E}[X]||_2 \leq ||X||_2$ (where $||X||_2^2 = {\rm Tr} X^\dagger X$). {This leads to the %following}
%\begin{Proposition} BLP-Markovianity implies
%\begin{equation}\label{}
%  \frac{d}{dt} ||\Delta_t \mathbf{x}||_2 \leq 0 ,
%\end{equation}
%for all $\mathbf{x} \in \mathbb{R}^{d^2-1}$.
%\end{Proposition}
%Equivalently, the purity ${\rm Tr}\rho^2$ satisfies $\frac{d}{dt} {\rm Tr}(\Lambda_t[\rho])^2 \leq 0$.

\vspace{.4cm}

{\em Hermitian commutative maps} --- {We conclude our theoretical analysis by considering} the most restrictive class of commutative maps, {namely those} satisfying $\Lambda_t^* = \Lambda_t$.
In this case $\lambda_\alpha(t)$ are real and since $\lambda_\alpha(0)=1$ and the map itself is invertible we have
%\begin{equation}\label{}
 $ \lambda_\alpha(t) = |\lambda_\alpha(t)|=s_\alpha(t)$.
%\end{equation}
In this case one has

\begin{Proposition}
If $\Lambda_t$ is a P-divisible commutative Hermitian dynamical map, then
%\begin{equation}
 $ B(t) \subset B(s)$,
%\end{equation}
for all $t \geq s$.
\end{Proposition}
Interestingly, in the case of Hermitian commutative maps we may provide an extra tool for analyzing P-divisibility (or equivalently BLP-Markovianity).
For a given dynamical map $\Lambda_t$ let us define the function
\begin{equation}\label{}
  f(t) = \<\alpha|(\oper \ot \Lambda_t)[P^+]|\alpha\> = d^{-2}\, {\rm Tr} F(t)\ .
\end{equation}
One has the following

\begin{Proposition}
If $\Lambda_t$ is a P-divisible commutative Hermitian map, then
\begin{equation}\label{f<0}
  \frac{d}{dt} f(t) \leq 0 ,
\end{equation}
for all $t \geq 0$.
\end{Proposition}
Indeed, since $\lambda_\alpha(0)=1$, then if $\lambda_\alpha(t)$ is real it must be positive otherwise it {would take null value} and the generator $\mathcal{L}_t$ {would become} singular. One has therefore
$$ \frac{d}{dt} \sum_\alpha \lambda_\alpha(t) =  \frac{d}{dt} \sum_\alpha |\lambda_\alpha(t)| \leq 0 , $$
due to (\ref{l}).  Note, that it is enough that $\Lambda_t$ is commutative and all eigenvalues are real (see Example - Amplitude damping channel).
We note here that condition (\ref{f<0}) can be easily detected in an
experimental scenario without performing all the measurements required for
quantum process tomography. Actually,
$f(t)$ is the probability of projecting
the global state of the system and the ancilla onto state $|\alpha\>$.
Let us consider for simplicity the case of two-dimensional systems.
We can write $|\alpha\>$ in terms of local Pauli operators as
$|\alpha\>\<\alpha|=1/4(I\otimes I+\sigma_x\otimes \sigma_x-\sigma_y\otimes
\sigma_y+\sigma_z\otimes \sigma_z)$. This means that $f(t)$ can be measured
from the expectation value of the local observables $\sigma_x\otimes \sigma_x,
\sigma_y\otimes \sigma_y,\sigma_z\otimes \sigma_z$ without requiring a complete
set of two-qubit operators that would be needed for entanglement-assisted
quantum process tomography. Moreover, the detection scheme considered here would be particularly suited
in a linear optical scenario. Actually, the projection onto the maximally
entangled state $|\alpha\>\<\alpha|$ could be performed in a single
measurement because it corresponds to a single projection onto a Bell state
while there is no need to distinguish between the four Bell states, which is
usually considered a drawback of linear optical implementations.

\vspace{.4cm}

{\em Examples of commutative dynamical maps}  --- Interestingly, this restricted class of maps -- commutative and normal/Hermitian -- still covers many interesting examples.
%Note that if $V_k^\dagger = V_k$, then
%\begin{eqnarray}\label{}
%  \mathcal{L}_t[\rho] &=& \frac 12 \sum_k \gamma_k(t) ( [V_k, \rho V_k] + [V_k\rho,V_k]) \nonumber \\ &=& -\frac 12 \sum_k \gamma_k(t) [V_k,[V_k,\rho]] ,
%\end{eqnarray}
%gives rise to Hermitian $\Lambda_t$. Now, we provide several examples of $V_k$ giving rise to commutative dynamical map.

\begin{Example}[Qubit dephasing] \label{EX-1} Consider
%\begin{equation}\label{s-z}
 $ \mathcal{L}_t[\rho] = \frac 12 \gamma(t)( \sigma_z \rho \sigma_z - \rho)$,
%\end{equation}
which gives rise to commutative Hermitian dynamical map. %One finds for eigenvalues of the corresponding dynamical map: $   \lambda_0(t)=\lambda_1(t)$ and  $\lambda_2(t)=\lambda_3(t) = e^{-\Gamma(t)}$,
%with $\Gamma(t) = \int_0^t \gamma(\tau) d\tau$.
For this very simple example all known conditions are equivalent: P-divisiblity (equivalently BLP-Markovianity) is equivalent to CP-divisibility, i.e., $\gamma(t) \geq 0$. In this case $B(t)$ defines an axially symmetric ellipsoid $ \frac{x_1^2}{\lambda^2(t)} + \frac{x_2^2}{\lambda^2(t)} + {x_3^2} \leq 1 $, where $\lambda(t) = e^{-\Gamma(t)}$ and $\Gamma(t) = \int_0^t \gamma(u)du$. Hence, this evolution is Markovian iff $B(t) \subset B(s)$ for $t > s$.
\end{Example}
This example may be generalized for $d>2$ {in two ways by providing normal or Hermitian time-local generator $\mathcal{L}_t$.} Note that $\{\sigma_0=\mathbb{I}/\sqrt{2},\sigma_1/\sqrt{2},\sigma_2/\sqrt{2},\sigma_3/\sqrt{2}\}$ define an orthonormal basis in $M_2(\mathbb{C})$ consisting of elements which are both Hermitian and unitary. Now, the unitary basis in $M_d(\mathbb{C})$ is defined by Weyl operators
%\begin{equation}\label{W}
$U_{kl}=\sum_{m=0}^{d-1} \omega^{mk}|m\>\<m+ l|$,
%\end{equation}
with $\omega=e^{2\pi i/d}$. If $d=2$ they reproduce four Pauli matrices.  Observe that $U_{k0}= \sum_{m=0}^{d-1} \omega^{mk}|m\>\<m|$ are diagonal and may be used to generalize Example \ref{EX-1}.

\begin{Example}[Qudit dephasing -- normal]\label{EX-2} Consider a qudit generator
\begin{equation}\label{}
  \mathcal{L}_t[\rho] =  \frac 12 \sum_{k=1}^{d-1}\gamma_k(t)( U_{k0} \rho U^\dagger_{k0} - \rho) .
\end{equation}
Clearly for $d=2$ one has $U_{10} = \sigma_z$ and the above generator reduces to that from Example~\ref{EX-1}. It is easy to check that $\mathcal{L}_t$ is normal and commutative.
\end{Example}
Consider now the Hermitian basis in $M_d(\mathbb{C})$ defined by Gell-Mann matrices. Diagonal elements are defined by
\begin{equation*}\label{}
  V_l = \frac{1}{\sqrt{l(l+1)}}\left( \sum_{k=0}^{l-1} |k\>\<k| - l |l\>\<l| \right) , \ \  l=1,\ldots,d-1 ,
\end{equation*}
and for $d=2$ one has $V_1 = \sigma_z /\sqrt{2}$.

\begin{Example}[Qudit dephasing -- Hermitian] \label{EX-3}
Consider a qudit generator
\begin{equation}\label{}
  \mathcal{L}_t[\rho] = -\frac 12 \sum_{l=1}^{d-1} \gamma_l(t) [V_l,[V_l,\rho]]  .
\end{equation}
It is easy to check that $\mathcal{L}_t$ is Hermitian and commutative. Diagonal elements $\rho_{kk}$ do not evolve in time and off-diagonal are multiplied by a function of local decoherence rates $\gamma_k(t)$.

% If $d=3$ one has two operators: $  V_1 = \frac{1}{\sqrt{2}} (|0\>\<0| - |1\>\<1|)$,
%and $  V_2 = \frac{1}{\sqrt{6}} (|0\>\<0| + |1\>\<1| - 2|2\>\<2|)$.
%The corresponding map $\Lambda_t$ has the following spectrum:  $\lambda_0(t) = \lambda_1(t)=\lambda_2(t)=1$,
%$ \lambda_3(t) = \lambda_4(t) = e^{-\Gamma_1(t)}$, and $  \lambda_5(t) = \ldots = \lambda_8(t) = e^{-\frac 14[\Gamma_1(t) + 3\Gamma_2(t)]}$.
%Conditions (\ref{l}) implies: $\gamma_1(t) \geq 0$ and $\gamma_1(t) + 3 \gamma_2(t) \geq 0$.
\end{Example}
%These two examples may be considered as a special cases of

\begin{Example}[Perfect decoherence -- normal]
Consider the following time-independent Hamiltonian in $\mathcal{H}_A \ot \mathcal{H}_B$
\begin{equation}\label{}
  H = H_A \ot \mathbb{I}_B + \mathbb{I}_A \ot H_B + \sum_k P_k \ot B_k ,
\end{equation}
where $P_k = |k\>\<k|$ are projectors into the computational basis vectors $|k\>$ in $\mathcal{H}_A$ and $B_k$ are hermitian operators in $\mathcal{H}_B$. Assuming that $H_A = \sum_k \epsilon_k P_k$ one finds
%\begin{equation}\label{}
 $ H = \sum_k P_k \ot Z_k$,
%\end{equation}
where $Z_k = \epsilon_k \mathbb{I}_B + H_B + B_k$. Such Hamiltonian leads to a pure decoherence of the density operator $\rho_A$ of subsystem A:
\begin{equation*}\label{}
  \rho_A(t) = {\rm tr}_B (e^{-iHt} \rho_A \ot \rho_B e^{iHt}) = \sum_{k,l} c_{kl}(t) P_k \rho_A P_l ,
\end{equation*}
with $c_{kl}(t) = {\rm tr}( e^{-i Z_k t} \rho_B e^{iZ_lt})$. It turns out that $c_{kl}(t)$ define eigenvalues of the map $\Lambda_t[\rho_A] = \sum_{k,l} c_{kl}(t) P_k \rho_A P_l$ which is commutative and normal.
%Note that Example \ref{EX-2} gives rise to the following matrix
%\begin{equation}\label{}
%  c_{kl}(t) = \exp\Big( \sum_{m=1}^{d-1} \Gamma_m(t) [ 1- \omega^{m(k-l)}] \Big) .
%\end{equation}
%BLP-Markovianity implies
%\begin{equation}\label{cos}
%  \sum_{m=1}^{d-1} \gamma_m(t) \Big[1-  \cos(2\pi m[k-l]/d)  \Big] \geq 0 ,
%\end{equation}
%for $k \neq l$. For $d=3$ one finds simple condition $\gamma_1(t) + \gamma_2(t) \geq 0$.
\end{Example}

\begin{Example}[Pauli channel -- Hermitian] The qubit dephasing  may be immediately generalized to
\begin{equation}\label{Pauli}
  \mathcal{L}_t[\rho] = \frac 12 \sum_{k=1}^3  \gamma_k(t)( \sigma_k \rho \sigma_k - \rho) ,
\end{equation}
which lead to the following dynamical map (time-dependent Pauli channel): $\Lambda_t[\rho] =  \sum_{\alpha=1}^3  p_\alpha(t) \sigma_\alpha \rho \sigma_\alpha $.  It was shown \cite{Filip-PRA} that (\ref{l}) implies: $ \gamma_1(t) +  \gamma_2(t) \geq 0$,  $\gamma_2(t) +  \gamma_3(t) \geq 0$, and  $\gamma_3(t) +  \gamma_1(t) \geq 0$. In this case $B(t)$ defines an ellipsoid $ \frac{x_1^2}{\lambda^2_1(t)} + \frac{x_2^2}{\lambda^2_2(t)} +\frac{x_3^2}{\lambda^2_3(t)} \leq 1 $ and this evolution is BLP-Markovian iff $B(t) \subset B(s)$ for $t > s$.
\end{Example}

\begin{Example}[Weyl channel -- normal] Pauli channel may be easily generalized for $d>2$ as follows
\begin{equation}\label{}
  \mathcal{L}_t[\rho] = \sum_{k+l>0}^{d-1} \gamma_{kl}(t)[ U_{kl} \rho U_{kl}^\dagger -\rho] ,
\end{equation}
where $U_{kl}$ are Weyl operators. This gives rise to the normal commutative dynamical map
$\Lambda_t[\rho] = \sum_{k,l=0}^{d-1} p_{kl}(t) U_{kl} \rho U_{kl}^\dagger$. Conditions (\ref{l}) lead to
$   \sum_{k+l>0} \gamma_{kl}(t) [1- {\rm Re}\omega^{mk - nl}] \geq 0  $
for all pairs $(m,n)$.
\end{Example}

\begin{Example}[Generalized Pauli channel]
Generalized Pauli channel \cite{Ruskai,KASIA} is a special example of the Weyl channel defined as follows: let $  \{ |\psi^{(\alpha)}_0\>,\ldots,|\psi^{(\alpha)}_{d-1}\> \}$ denote $d+1$  mutually unbiased bases (MUBs) in $\mathbb{C}^d$. Define the quantum channels $\mathcal{P}_\alpha[\rho] = \sum_{l=0}^{d-1}  | \psi^{(\alpha)}_l\> \<  \psi^{(\alpha)}_l| \rho |  \psi^{(\alpha)}_l\>\<\psi^{(\alpha)}_l|$ and let
\begin{equation}\label{}
 \mathcal{L}_t[\rho]= \sum_{\alpha=1}^{d+1} \gamma_\alpha(t) \left( \mathcal{P}_\alpha[\rho] - \rho \right) ,
  %\Lambda_{t}[\rho] = p_0(t)  \rho + \frac{1}{d-1}\sum_{\alpha=1}^{d+1} p_\alpha(t) \mathbb{U}_\alpha[\rho] ,
\end{equation}
 This map is Hermitian and BLP-Markovianity implies \cite{KASIA} $
%\begin{equation}\label{}
  \gamma(t) - \gamma_\alpha(t) \geq 0$,
%\end{equation}
where $\gamma(t) = \sum_\alpha \gamma_\alpha(t)$.

\end{Example}

\begin{Example}[Amplitude damping channel]
The dynamics of a single amplitude-damped qubit is
governed by a single function $G(t)$
\begin{equation}\label{}
  \Lambda_t[\rho] = \left( \begin{array}{cc} \rho_{11} + (1-|G(t)|^2)\rho_{22} & G(t) \rho_{12} \\ G^*(t) \rho_{21} & |G(t)|^2 \rho_{22} \end{array} \right) ,
\end{equation}
where the function $G(t)$ depends on the form of the reservoir spectral density $J(\omega)$ \cite{Breuer}. The dynamical map $\Lambda_t$ is commutative but not normal. The corresponding eigenvalues read as follows: $\lambda_0(t)=1$, $\lambda_1(t)=G(t)$, $\lambda_2(t)= G^*(t)$, and $\lambda_3(t) = |G(t)|^2$.
This evolution is generated by the following time-local generator
\begin{equation*}\label{}
  \mathcal{L}_t[\rho] = - \frac{is(t)}{2}[\sigma_+\sigma_-,\rho] + \gamma(t)( \sigma_- \rho_+ - \frac 12 \{ \sigma_+\sigma_-,\rho\} ) ,
\end{equation*}
where $\sigma_\pm$ are the spin lowering and rising operators together with
$s(t) = -2{\rm Im} \frac{\dot{G}(t)}{G(t)}$, and $\gamma(t) = -2{\rm Re} \frac{\dot{G}(t)}{G(t)}$. It is clear that (\ref{l}) implies $\gamma(t) \geq 0$. Again in this case this condition is necessary and sufficient for Markovianity. Since eigenvalues are in general complex we cannot use condition (\ref{f<0}).
Interestingly, for Lorentzian spectral density $J(\omega) = \frac{\gamma_M \lambda^2}{2\pi[ (\omega-\omega_c)^2 + \lambda^2]}$ the function $G(t)$ becomes real and hence $ f(t) = \frac 14[1 + G(t)]^2 $ and condition (\ref{f<0}) implies $\gamma(t) \geq 0$. This example may be considered as an analog of non-Hermitian Hamiltonian with real spectra analyzed by Bender \cite{Bender}. 
%Note, however, that it is well known that for Lorentzian spectral density the evolution is always Markovian.

 %\RED{COMMENT: ABOUT THIS EXAMPLE, I THINK THAT G(t) IS REAL WHEN THE TWO-STATE SYSTEM IS RESONANT WITH THE PEAK OF THE LORENTZIAN, HOWEVER IN THIS CASE THE %DYNAMICAL MAP IS NOT INVERTIBLE FOR VALUES OF THE PARAMETERS IN WHICH IT'S NON-MARKOVIAN.}

%\blue{But in this case one finds

%$$  G(t) = e^{-\lambda t/2} \Big( \cosh(\Omega t/2) + \frac{\lambda}{\Omega} \sinh (\Omega t/2) \Big)$$,
%where $\Omega= \sqrt{\lambda^2 - 2 \lambda \gamma_M}$. Hence, if $\lambda^2 - 2 \lambda \gamma_M \geq 0$ one has $G(t) > 0$ which means that the map is %invertible (all eigenvalues are strictly positive)}.
\end{Example}

\vspace{.1cm}

{\em Conclusions} --- In this Letter we provided further characterization of non-Markovian evolution for a class of commutative dynamical maps.
In this case P-divisibility implies simple conditions for the spectrum of the  dynamical map. Moreover, if the map is normal then P-divisibility is equivalent to BLP-Markovianity and the body of accessible states $B(t)$ is contained up to orthogonal rotation in $B(s)$ for $t > s$. This provides a much stronger non-Markovianity witness {than the}  volume of accessible states \cite{Pater}. Finally, it is argued that the quantity $ \<\alpha|(\oper \ot \mathcal{L}_t)[P^+]|\alpha\>$ may be considered as a  dynamical analog of entanglement witness that can be easily accessed in the experimental scenario. Our analysis is illustrated by several {paradigmatic} examples.

\vspace{.4cm}

%\acknowledgements

{\em Acknowledgements} --- DC was partially supported by the National Science Centre project 2015/17/B/ST2/02026. {S.M. acknowledges financial support from the EU project QuProCS (Grant Agreement 641277), the Academy of Finland (Project no. 287750) and the Magnus Ehrnrooth Foundation.}

%\begin{widetext}

\section*{Supplementary Material}

Any positive trace-preserving map is a contraction in the trace norm
$$   ||\Phi[X]||_1 \leq ||X||_1 , $$
where $||X||_1 = {\rm Tr} |A| = {\rm Tr}\sqrt{XX^\dagger}$. Now, if $\Phi$ is not only trace-preserving but also unital (doubly stochastic), then it is also a contraction in the Hilbert-Schmidt norm
$$   ||\Phi[X]||_2 \leq ||X||_2 , $$
for all normal $X$. Recall that $||X||_2 = {\rm Tr}{XX^\dagger}$. Indeed, using the Kadison inequality,
\begin{equation*}
  \Phi[X^\dagger X] \geq \Phi[X^\dagger] \Phi[X] ,
\end{equation*}
one has
\begin{eqnarray*}
  || \Phi[X]||_2^2 &=& {\rm Tr}(\Phi[X^\dagger] \Phi[X]) \leq {\rm Tr}(\Phi[X^\dagger X]) \nonumber \\ &=& {\rm Tr}(X^\dagger X) = ||X||_2^2 .
\end{eqnarray*}
Suppose now that $\Lambda_t$ is P-divisible and unital. One has for any normal $X$
\begin{eqnarray*}\label{}
   \frac{d}{dt} ||\Lambda_t[X] ||_2 &=& \lim_{\epsilon \rightarrow 0} \frac 1\epsilon \Big( ||\Lambda_{t+\epsilon}[X]||_2 - ||\Lambda_t[X]||_2 \Big) \\
   &=& \lim_{\epsilon \rightarrow 0} \frac 1\epsilon \Big( ||V_{t+\epsilon,t}[\Lambda_t[X]]||_2 - ||\Lambda_t[X]||_2 \Big) \\
   &\leq& \lim_{\epsilon \rightarrow 0} \frac 1\epsilon \Big( ||\Lambda_t[X]||_2 - ||\Lambda_t[X]||_2 \Big) = 0 ,
\end{eqnarray*}
where we used the fact that $V_{t+\epsilon,t}$ is a contraction in Hilbert-Schmidt norm.

To prove
$$    \frac{d}{dt} || \Delta_t \mathbf{x} ||_2 \leq 0 , $$
let us consider a Hermitian (and hence normal) operator
$$  X = x_0 \mathbb{I} + \sum_{k} x_k G_k . $$
One has
\begin{eqnarray*}\label{}
   && ||\Lambda_t[X] ||^2_2 ={\rm Tr}\Big[\Big(x_0 \mathbb{I} + \sum_{k} x_k \Lambda_t[G_k]\Big)\Big(x_0 \mathbb{I} + \sum_{l} x_l \Lambda_t[G_l]\Big)\Big] \\
    && =x_0^2 d + 2 x_0 \sum_{k} x_k  {\rm Tr}(\Lambda_t[G_k]) + \sum_{k,l} x_k x_l  {\rm Tr}(\Lambda_t[G_k]\Lambda_t[G_l]) .
\end{eqnarray*}
Now, since Gell-Mann matrices are traceless one has ${\rm Tr}(\Lambda_t[G_k]) = {\rm Tr}G_k = 0$. Moreover, using
$$  \Lambda_t[G_k] = \sum_m \Delta_{km}(t) G_m  , $$
one finds
$$  \sum_{k,l} x_k x_l  {\rm Tr}(\Lambda_t[G_k]\Lambda_t[G_l]) =   \sum_{k,l,m} x_k x_l \Delta_{km}(t)  {\rm Tr}(G_m\Lambda_t[G_l]) $$
and using
$$  \Delta_{ml}(t) = {\rm Tr}(G_m\Lambda_t[G_l]) $$
one arrives to
$$  \sum_{k,l} x_k x_l  {\rm Tr}(\Lambda_t[G_k]\Lambda_t[G_l]) =  \sum_{k,l,m} x_k x_l \Delta_{km}(t) \Delta_{ml}(t) $$
and finally
$$  ||\Lambda_t[X] ||^2_2 =  x_0^2 d + ||\Delta_t \mathbf{x}||^2_2 , $$
where now $||\mathbf{x}||_2^2 = \sum_k x_k^2$. It is, therefore, clear that
$$     \frac{d}{dt} || \Delta_t \mathbf{x} ||_2 =  \frac{d}{dt} ||\Lambda_t[X] ||_2 \leq 0 . $$


\begin{thebibliography}{1} \bibliographystyle{plain}

\bibitem{VN} J. von Neumann, {\em Mathematical Foundations of Quantum Mechanics},

\bibitem{Breuer} H.-P. Breuer and F. Petruccione,
{\em The Theory of Open Quantum Systems} (Oxford Univ. Press,
Oxford, 2007).

\bibitem{Weiss} U. Weiss, {\it Quantum Dissipative Systems}, (World
Scientific, Singapore, 2000).

\bibitem{Alicki} R. Alicki and K. Lendi, {\it Quantum Dynamical
Semigroups and Applications} (Springer, Berlin, 1987).

\bibitem{QIT} M. A. Nielsen and I. L. Chuang, {\it Quantum
Computation and Quantum Information} (Cambridge Univ. Press,
Cambridge, 2000).

%======================

\bibitem{rev1}\'A. Rivas, et al., Rep. Prog. Phys. {\bf77}, 094001 (2014).
%S.F. Huelga, and M.B. Plenio, Rep. Prog. Phys. {\bf77}, 094001 (2014).

\bibitem{rev2}H. -P. Breuer, et al.,
%E. -M. Laine, J. Piilo, B. Vacchini, Rev. Mod.
Phys. {\bf 88}, 021002 (2016).

%========================
\bibitem{HHHH} R. Horodecki, et al., Rev. Mod. Phys. \textbf{81}, 865 (2009).

% P. Horodecki, M. Horodecki and K. Horodecki, {\em Quantum entanglement}, Rev. Mod. Phys. \textbf{81}, 865 (2009).


%\bibitem{R1}  H.-P. Breuer and B. Vacchini, Phys.
%Rev. Lett. {\bf 101} (2008) 140402; Phys. Rev. E {\bf 79}, 041147
%(2009).

%\bibitem{R2}  J. Piilo, S. Maniscalco, K. H\"ark\"onen and K.-A. Suominen, Phys. Rev. Lett. {\bf 100}, 180402
%(2008); Phys. Rev. A {\bf 79}, 062112 (2009).

%\bibitem{R3} W.-M. Zhang, P.-Y. Lo, H.-N. Xiong, M. W.-Y. Tu, and F. Nori, Phys. Rev. Lett. {\bf 109}, 170402 (2012).

%\bibitem{R4} D. Chru\'sci\'nski and A. Kossakowski, Phys. Rev. Lett. {\bf
%104}, 070406 (2010); Phys. Rev. Lett. {\bf 111}, 050402 (2013).



%\bibitem{Saverio} D. Chru\'sci\'nski, A. Kossakowski, and S.
%Pascazio, Phys. Rev. A {\bf 81}, 032101 (2010).





\bibitem{Wolf-Isert} M. M. Wolf, et al., %J. Eisert, T. S. Cubitt and J. I. Cirac,
Phys. Rev. Lett. \textbf{101}, 150402 (2008).

\bibitem{SM} Supplementary Material

\bibitem{RHP} \'A. Rivas, et al., %S.F. Huelga, and M.B. Plenio,
Phys. Rev. Lett. {\bf 105}, 050403 (2010).


%\bibitem{Hou} S. C. Hou, X. X. Yi, S. X. Yu, and C. H. Oh, Phys. Rev. A {\bf 83},
%062115 (2011).

\bibitem{BLP} H.-P. Breuer, et al., %E.-M. Laine, J. Piilo,
Phys. Rev. Lett. {\bf 103}, 210401 (2009).

%======================

\bibitem{Fisher} X.-M. Lu, et al., % X. Wang, and C. P. Sun,
Phys. Rev. A {\bf 82}, 042103 (2010).

\bibitem{fidelity} A. K. Rajagopal, et al., %A. R. Usha Devi, and R. W. Rendell,
Phys. Rev. A 82, 042107 (2010).

\bibitem{Luo1} S. Luo, et al.,  Phys. Rev. A 86, 044101
(2012).

%S. Fu, and H. Song, Phys. Rev. A 86, 044101
%(2012).

\bibitem{Luo2} M. Jiang and S. Luo,  Phys. Rev. A  {\bf 88}, 034101 (2013).


\bibitem{Bogna}  B. Bylicka, et al., Scientific Reports, {\bf 4}, 5720 (2014).

\bibitem{Pater}  S. Lorenzo, et al., Phys. Rev. A {\bf 88}, 020102(R) (2013).

\bibitem{WOLF}  M. Wolf and J. I. Cirac, Commun. Math. Phys. {\bf 279}, 147168 (2008).

%F. Plastina, M. Paternostro, Phys. Rev. A {\bf 88}, 020102(R) (2013).

\bibitem{Adesso} A.M. Leonardo, et al., Phys. Rev. A {\bf 91}, 032115 (2015).
%======================

\bibitem{MR}
C. Macchiavello and M. Rossi, Phys. Rev. A  {\bf 88}, 042335 (2013).

\bibitem{exp-qcd}
A. Orieux, et al. Phys. Rev. Lett. {\bf 111}, 220501 (2013).


\bibitem{PRL-Sabrina}  D. Chru\'sci\'nski and S. Maniscalco, Phys. Rev. Lett, {\bf 112}, 120404 (2014).

\bibitem{test} N.K. Bernardes, et al., Scientific Reports {\bf 5}, 17520 (2015)

%\bibitem{RECENT} F. Benatti, R. Floreanini, and G. Scholes eds., {\it Special issue on loss of coherence and memory effects in quantum dynamics},  J. Phys. B {\bf 45}, No. 15 (2012).

%========================

%\bibitem{optimal} S. Wissmann, A. Karlsson, E.-M. Laine, J. Piilo, and H.-P. Breuer,   {\em   Optimal state pairs for non-Markovian quantum dynamics}, %arXiv:1209.4989.


%\bibitem{GKS} V. Gorini, A. Kossakowski, and E. C. G. Sudarshan, J. Math. Phys.
%{\bf 17}, 821 (1976).

%\bibitem{Lindblad}  G. Lindblad, Comm. Math. Phys. {\bf 48}, 119
%(1976).

%===========

%\bibitem{versus1} P. Haikka, J. D. Cresser, and S. Maniscalco,  Phys. Rev. A {\bf 83}, 012112 (2011).

%\bibitem{versus2} B. Vacchini, A. Smirne, E.-M. Laine, J. Piilo, and H.-P. Breuer,
%New J. Phys. {\bf 13}, 093004 (2011).

%\bibitem{versus3} D. Chru\'sci\'nski, Kossakowski and \'A. Rivas,  Phys. Rev. A {\bf 83}, 052128 (2011).


%===================

%\bibitem{HHHH} R. Horodecki, P. Horodecki, M. Horodecki and K. Horodecki, Rev. Mod. Phys. \textbf{81}, 865 (2009).


%\bibitem{Pawel} P. Horodecki and B. Terhal,  Phys. Rev. A {\bf 61}, 040301(R) (2000).


%========non-M=================


%\bibitem{JC} D. Maldonado-Mundo, P. Ohberg, B. W. Lovett, E. Andersson,  Phys. Rev. A {\bf 86}, 042107 (2012).

%\bibitem{nM2} A. Shabani and D.A. Lidar, Phys. Rev. A {\bf 71}, 020101(R) (2005).



%\bibitem{Sanpera} A. Sanpera, D.  Bruss, and M. Lewenstein, Phys. Rev. A {\bf 63}, 050301 (2001).



%\bibitem{EPL} D. Chru\'sci\'nski and A. Kossakowski, EPL, {\bf 97}, 20005 (2012).

%\bibitem{PRL-2013} D. Chru\'sci\'nski and A. Kossakowski, Phys. Rev. Lett. {\bf
%111}, 050402 (2013).

%\bibitem{Franco-PRA} R. Lo Franco, B. Bellomo, E. Andersson, and G. Compagno,  Phys. Rev. A {\bf 85}, 032318 (2012).

 %M. Mannone, R. Lo Franco, and G. Compagno, {\em Comparison of non-Markovianity criteria in a qubit system under random external fields}, %Physica Scripta T, in press (2012) [arXiv:1209.6331].

%\bibitem{JPB} D. Chru\'sci\'nski and A. Kossakowski,  J. Phys. B: At. Mol. Opt. Phys. {\bf 45},  154002 (2012).



%===================measure=non-Markovian============================



%=================inne miary=============

%\bibitem{inne1} X.-M. Lu, X. Wang, and C. P. Sun, Phys. Rev. A {\bf 82}, 042103
%(2010).

%\bibitem{inne2} A. K. Rajagopal, A. R. Usha Devi, and R. W. Rendell, Phys.
%Rev. A {\bf 82}, 042107 (2010).

%\bibitem{inne3} S. C. Hou, X. X. Yi, S. X. Yu, and C. H. Oh, Phys. Rev. A {\bf 83},
%062115 (2011).

%\bibitem{inne3a} S. C. Hou, X. X. Yi, S. X. Yu, and C. H. Oh, Phys. Rev. A {\bf 86}, 012101 (2012).

%\bibitem{inne4} S. Luo, S. Fu, and H. Song, Phys. Rev. A {\bf 86}, 044101 (2012).

%\bibitem{Gauss} R. Vasile, S. Maniscalco, M. G. A. Paris, H.-P. Breuer, and
%J. Piilo, Phys. Rev. A {\bf 84}, 052118 (2011).

%==================================================

%\bibitem{inne}


%\bibitem{inne1} E.-M. Laine, J. Piilo, and H.-P. Breuer, Phys. Rev. A {\bf 81}, 062115
%(2010).

%\bibitem{inne2} L. Mazzola, E.-M. Laine, H.-P. Breuer, S. Maniscalco, and J.
%Piilo, Phys. Rev. A {\bf 81}, 062120 (2010).

%\bibitem{Apollaro} T. J. G. Apollaro, C. Di Franco, F. Plastina, and M.
%Paternostro, Phys. Rev. A {\bf 83}, 032103 (2011).

%\bibitem{Hu} C.H. Fleming, B.L. Hu, Ann. Phys. (N.Y.), {\bf 327},  1238 (2012).

%\bibitem{Benatti} F. Benatti, R. Floreanini, and S. Olivares, Phys. Lett. A {\bf 376}, 2951 (2012).

%\bibitem{B-review} H.-P. Breuer, J. Phys. B: At. Mol. Opt. Phys. {\bf 45}, 154001 (2012).

%\bibitem{Buzek}  P. Pechukas, Phys. Rev. Lett. {\bf 73}, 1060 (1994); R. Alicki, Phys. Rev. Lett. {\bf 75}, 3020 (1995); P. Stelmachovic and V. Buzek, Phys. Rev. A {\bf 64},  062106 (2001);  A. Shaji and E.C.G. Sudarshan, Phys. Lett. A {\bf 341}, 48 (2005).

%\bibitem{Paulsen} V. Pauslen, {\it Completely Bounded Maps and Operator Algebras}, Cambridge University Press, 2002.

%\bibitem{PLA} D. Chru\'sci\'nski and F. Wudarski, Phys. Lett. A {\bf 377}, 1425 (2013).

\bibitem{Filip-PRA} D. Chru\'sci\'nski, F. A. Wudarski,  Phys. Rev. A {\bf 91}, 012104 (2015).


%\bibitem{Petz} M. Ohya and D. Petz, {\em Quantum Entropy and Its Use}, Springer, Berlin, 2004.

%\bibitem{apollaro} T.J.G. Apollaro, {\it et al.}, {\em Competition between memory-keeping and memory-erasing decoherence channels}, arXiv:1311.2045.

%\bibitem{Wolf} M.~M. Wolf and J.~I. Cirac, Comm. Math. Phys. \textbf{279}, 147 (2008).


%\bibitem{opt} M. Lewenstein, {\it et al.}, Phys. Rev A. {\bf 62}, 052310 (2000).

\bibitem{damping} H.-J.  Briegel and B.-G. Englert, Phys. Rev. A {\bf 47}, 3311  (1993).

\bibitem{Erika} E. Andersson, et al., %J. D. Cresser, and M. J. W. Hall,
Phys. Rev. A {\bf 89}, 042120 (2014).


\bibitem{Ruskai}  M. Nathanson and M.B Ruskai, J. Phys. A: Math. Theor. {\bf 40} 8171 (2007).

\bibitem{KASIA} D. Chru\'sci\'nski and K. Siudzi\'nska,  arXiv:1606.02616.

\bibitem{Bender}  C. M. Bender and S.  Boettcher,   Phys. Rev. Lett. {\bf 80}, 5243 (1998).



\end{thebibliography}
\end{document}